\documentclass[epj]{svjour}

\usepackage{graphicx}
\begin{document}

\title{Isoscalar dipole coherence at low energies and forbidden E1 strength} 

\author{P.~Papakonstantinou\thanks{Email: panagiota.papakonstantinou@physik.tu-darmstadt.de} 
\and  
V.Yu.~Ponomarev 
\and 
R.~Roth 
\and 
J.~Wambach} 
\institute{Institut f\"ur Kernphysik, 
Technische Universit\"at Darmstadt, 
Schlossgartenstr.~9, 
D-64289 Darmstadt, Germany}

\abstract{ 
In 
$^{16}$O and $^{40}$Ca an isoscalar, low-energy dipole transition (IS-LED) exhausting approximately $4\%$ of the isoscalar dipole (ISD) 
energy-weighted sum rule is experimentally known, but conspicuously absent from recent theoretical investigations of ISD strength. 
The IS-LED mode coincides with the so-called isospin-forbidden $E1$ transition. 
We report that 
for $N=Z$ nuclei up to $^{100}$Sn 
the fully self-consistent Random-Phase-Approximation 
with finite-range forces, phenomenological and realistic, 
yields a collective IS-LED mode, typically overestimating its excitation energy, but 
correctly describing its IS strength and electroexcitation form factor. 
The presence of $E1$ strength is solely due to the Coulomb interaction between the protons and the resulting isospin-symmetry breaking.  
The smallness of its value is related to the form of the transition density, due to translational invariance.  
The calculated values of $E1$ and ISD strength carried by the IS-LED depend on the effective interaction used. 
Attention is drawn to the possibility that in $N\neq Z$ nuclei this distinct mode of IS surface vibration can develop as such  
or mix strongly with skin modes and thus influence the pygmy dipole strength as well as the ISD strength function. 
In general, theoretical models currently in use may be unfit to predict its precise position and strength, if at all its existence.  
} 


 


\maketitle

\section{Introduction} 
\label{S:intro} 

It is an experimental observation that nuclei undergo iso\-sca\-lar dipole (ISD) transitions in the $1\hbar\omega$ regime of excitation energy~\cite{HaV2001s1}. 
The low-energy ISD strength function in $N>Z$ nuclei has received much attention, either indirectly in the context 
of neutron-skin modes or in the form of toroidal modes~\cite{PVK2007}. 
Absent from recent theoretical investigations are the IS low-energy dipole (IS-LED) modes of $N=Z$ nuclei, 
notably $^{16}$O and $^{40}$Ca, where the $J^{\pi}=1^-$, $T=0$ states at $7.12$~MeV and $6.95$~MeV, respectively, 
exhaust approximately $4\%$ of the ISD energy-weighted sum rule (EWSR)~\cite{HaD1981,Poe1992} 
and carry little, but not negligible, $E1$ strength. 
Note, for example, that the $6.95$~MeV state of $^{40}$Ca  carries practically all pygmy dipole strength below $10$~MeV in this nucleus~\cite{Har2004}  
(a weaker $1^-$ transition at $5.90$~MeV has been attributed to a rotational band~\cite{Roe2004}). 

The electroexcitation of the above states reveals a diffraction minimum in the longitudinal form factors 
\cite{MiS1975,FrV1989,Gra1977}. 
In the mid-seventies and for some years thereafter 
much theoretical effort was directed at accounting for their $E1$ strength, as a measure of isospin-symmetry violation, and form factors. 
A weak isospin mixing in the ground or excited state could explain the findings, though  
the numerical results were very sensitive to the input, including the isospin mixing assumed~\cite{AMS1975,Bar1976,HeS1976,COS1990}. 
Only a scant few explicit reports exist on these states within the self-consistent Random-Phase Approximation (RPA), 
mostly regarding their position and isospin content~\cite{BlG1977,Hey1990}.  

In this work we present self-consistent RPA calculations of the IS-LED states in $N=Z$ spherical, closed-shell nuclei. 
We focus on such nuclei, because of the detailed experimental data available for $^{16}$O and $^{40}$Ca 
and because the IS-LED states are found less fragmented. 
Thus an unambiguous comparison with the experimental spectrum is possible. 
We use finite-range interactions, both phenomenological (Gogny and Brink-Boeker) and realistic or semi-realistic (unitarily transformed AV18, 
with or without a phenomenological three-body term). 
In self-consistent Hartree-Fock--RPA (HF-RPA), such as employed here,  
the Coulomb interaction and resulting isospin mixing is either included in both the ground and the excited states or ignored altogether, 
unlike, e.g., the valence shell model, where a separate treatment of the $ph$ energies and wavefunctions and the $ph$ interaction is customary. 
We will find that the IS-LED state in all studied nuclei is a collective $1\hbar\omega$ transition. 
The interactions we use tend to overestimate the excitation energy and some the $E1$ strength of the IS-LED state, 
but the overall good comparison with the available data
suggests that the model accounts for the correct physics. 
The presence of $E1$ strength and the smallness of its value, despite the collectivity of the state, 
are elegantly explained as due to the Coulomb interaction and translational-invariance requirements, respectively. 
We attempt predictions for the IS-LED states in the unstable nuclei $^{56}$Ni and $^{100}$Sn. 

As we will discuss, there are indications that other RPA models do not produce any prominent IS-LED. 
This may be due to not fully self-consistent calculations. 
At any rate, no effective interaction or microscopic model, in general, has been tuned to describe the IS-LED properties 
and few have been tried at all. 
Thus the following question arises: can theoretical models currently in broad use describe correctly the low-energy ISD and $E1$ strength 
in stable, experimentally well-studied nuclei, and if not, to what extent can they be relied upon to describe the properties of exotic dipole modes in, 
e.g., very neutron-rich species.   
Although we will not attempt an answer to the second part of the question here, 
we regard the present exploratory work as an initiative to start examining such critical matters.

In Sec.~\ref{S:theory} we present the formalism used in this work. 
In Sec.~\ref{S:res}, results on the $1^-$ response of selected nuclei are presented 
and the properties of the IS-LED are analyzed and compared with experimental measurements. 
In Sec.~\ref{S:disc} they are discussed 
in the broader context of effective interactions and of pygmy dipole strength. 
We summarize in Sec.~\ref{S:concl}.

\section{Theory} 
\label{S:theory} 

We employ the self-consistent HF-RPA for closed-shell nuclei. 
The HF problem is solved within a single-particle basis spanning $13-15$ harmonic-oscillator shells. 
The same effective interaction is used to construct the RPA equations, solved within the HF basis. 
In particular, we employ a two-body Hamiltonian of the form 
\begin{equation} 
H = T + V_{\mathrm{NN}} + V_{\mathrm{Coul}} + V_{\rho} 
, 
\end{equation} 
where $T$ is the intrinsic kinetic energy, $V_{\mathrm{NN}}$ a nucleon-nucleon interaction excluding the Coulomb term $V_{\mathrm{Coul}}$ 
acting between protons and 
\begin{equation} 
V_{\rho} = t_3(1+x_3)\delta (\vec{r})\rho^{\alpha}(\vec{R}) 
\end{equation} 
is a density-dependent contact interaction ($\vec{r}$ the relative and $\vec{R}$ the center-of-mass position vector of the interacting nucleon pair). 
For $\alpha=1$, $V_{\rho}$ is equivalent to a three-body contact interaction. 

We employ various finite-range NN interactions, both phenomenological and realistic ones. 
We will present results mainly with the Gogny D1S~\cite{BGG1991} parameterization 
and a unitarily-transformed AV18 realistic potential, supplemented with a phenomenological three-body contact term~\cite{GRH2010}, 
which we will label here UCOM(SRG)$_{\mathrm{S,\delta 3N}}$. 
The latter is determined by transforming the $S-$waves of AV18 using the unitary correlation operator method (UCOM) 
with correlation functions determined via the similarity renormalization group (SRG).
We also use 
another AV18-based potential, which we will label SRG$_{\mathrm{S,\delta 3N}}$, resulting from transforming the $S-$wave channels of AV18 
using SRG~\cite{GRH2010}. 
The three-body term is determined in both 
UCOM(SRG)$_{\mathrm{S,\delta 3N}}$ and SRG$_{\mathrm{S,\delta 3N}}$ 
such that ground-state properties are well reproduced throughout the nuclear 
chart within perturbation theory ($C_{3N}=6t_3=2200$ and $2000$~MeV~fm$^6$, respectively, $x_3=1.0$ in both cases).  
The flow parameters are 0.16~fm$^4$ and 0.10~fm$^4$, respectively. 
For comparison,  
the pure two-body UCOM-transformed AV18, UCOM$_{\mathrm{var}}$, already employed in several studies, e.g., \cite{RPP2006,PaR2009}, 
is adopted as well. 
Regarding the UCOM and SRG procedures and  the properties of the various interactions we refer the reader to ref.~\cite{RNF2010}.  
Finally, 
for the $\ell -$closed nuclei $^{16}$O and $^{40}$Ca  
we have used also the two-body central Brink-Boeker B1 interaction.

The ISD response is calculated for the transition operator 
\begin{equation} 
\hat{O}_{\mathrm{ISD}} = \sum_{i=1}^A (r_i^3 - \frac{5}{3}\langle r^2\rangle r_i)Y_{1m}(\Omega_i )
\end{equation} 
and the electromagnetic response using 
\begin{equation} 
\hat{O}_{E1} = \frac{Z}{A}\sum_{n=1}^N r_n Y_{1m}(\Omega_n ) - \frac{N}{A}\sum_{p=1}^{Z} r_p Y_{1m}(\Omega_p ) 
\end{equation} 
in an obvious notation, where the subscripts $p$ and $n$ refer to protons and neutrons, respectively. 
We calculate the excitation strength, $B(E1\uparrow )$. 
The above operators include corrections to explicitly restore translational invariance. 
However, because our calculations are fully self-consistent, we obtain practically the same values of strength 
if we use the uncorrected forms of these operators, 
\begin{equation} 
\hat{O}_{\mathrm{ISD}}^{(0)} = \sum_{i=1}^A r_i^3Y_{1m}(\Omega_i )   
\quad , \quad 
\hat{O}_{E1}^{(0)} = \sum_{p=1}^{Z} r_p Y_{1m}(\Omega_p ) ,  
\end{equation} 
except of course for the spurious state, which appears at practically zero energy. 

Electroexcitation cross sections  
are calculated by using the proton transition density, $\delta\rho_p(r)$, 
\begin{equation} 
\delta\rho (r)= \delta\rho_p(r) .
\label{E:noeffch}
\end{equation} 
We warn against the use of a corrected isovector (IV) transition density instead, using effective charges as in 
\begin{equation} 
\delta\rho_{\mathrm{IV}} (r)= \frac{Z}{A} \delta\rho_n(r) - \frac{N}{A} \delta\rho_p(r) .
\label{E:effch}
\end{equation} 
The two procedures yield very different results, perhaps coinciding close to the photon point in self-consistent calculations. 
The reason is that effective charges as above have been derived and are only applicable for use in the long-wavelength limit. 
It is of course important to remove the spurious admixtures, if any, from the wavefunction or transition density before calculating 
the form factor -- or at least use a different prescription for testing and effectively correcting the form factor~\cite{Pap2005}. 

For the longitudinal form factors in plane-wave Born approximation (PWBA) we use the convention 
\begin{equation} 
F_1(q^2) = \frac{\sqrt{12\pi}}{Z} \int_0^{\infty} \delta\rho_p (r) j_1(qr)r^2 \mathrm{d} r 
.
\end{equation} 
Note that some results reported in the literature may be lacking factors $1/Z$ or $\sqrt{4\pi}$, or other. 
Eventually, the cross section divided by the Mott cross section takes the place of the form factor squared, in distorted-wave Born approximation (DWBA). 

\section{Results} 
 \label{S:res} 
 
\subsection{General features of the dipole response} 

\label{sS:general} 

\begin{figure*} 
\centering\includegraphics[angle=-90,width=15.0cm]{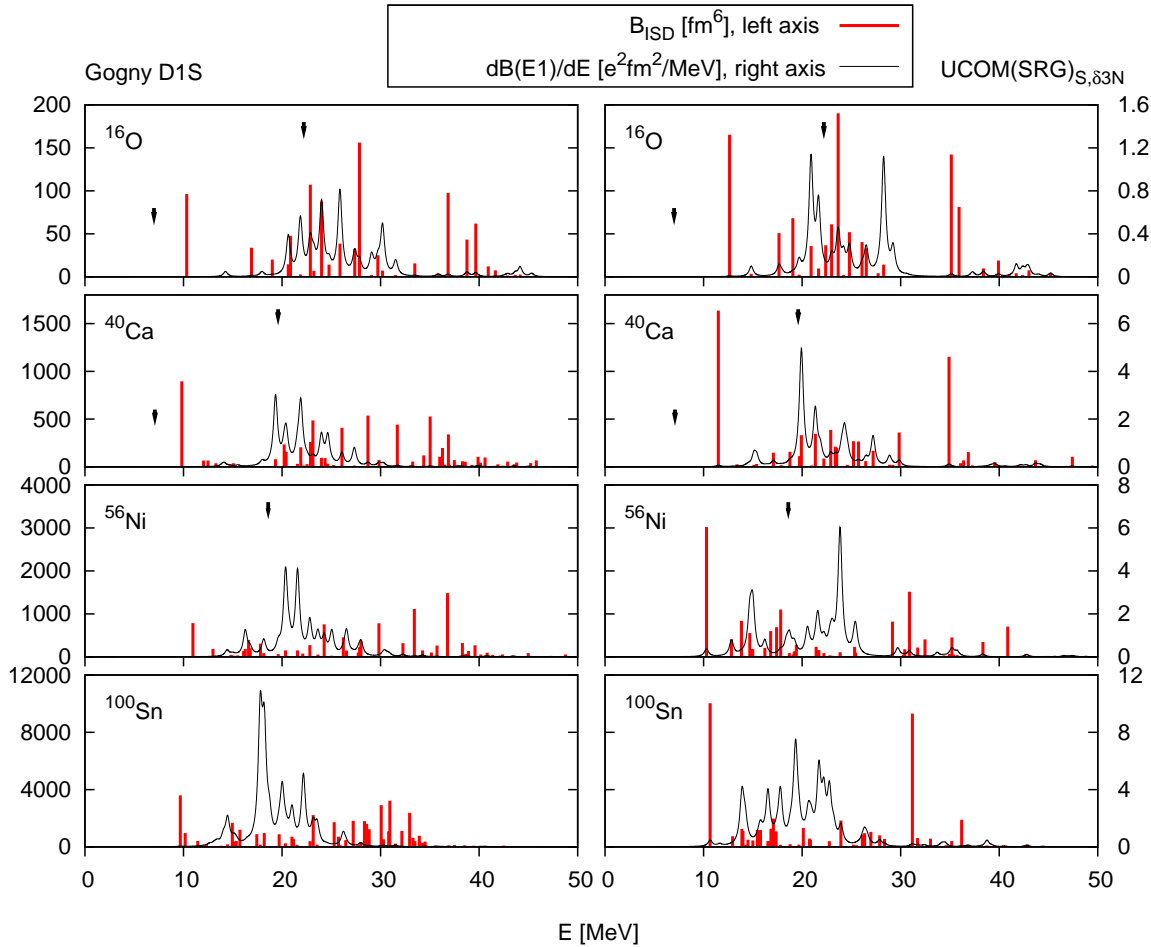}\\[2mm]  
\caption{%
ISD and $E1$ response of the $Z=N$ nuclei $^{16}$O, $^{40}$Ca, $^{56}$Ni, and $^{100}$Sn, 
within self-consistent HF-RPA using the interactions Gogny D1S and UCOM(SRG)$_{S,\mathrm{\delta 3N}}$, a transformed AV18 plus three-body term (see text). 
The $E1$ strength function has been folded with a $0.5$~MeV-wide Lorenzian, for visibility. 
Arrows mark the position of the main GDR peak (from photoabsorption cross section data~\cite{DiB1988}; for $^{56}$Ni the datum for $^{58}$Ni is given) and, 
for $^{16}$O and $^{40}$Ca, also the IS-LED.  
\label{F:response}}
\end{figure*}

In fig.~\ref{F:response} we show the ISD and $E1$ strength functions of the four $N=Z$ nuclei $^{16}$O, $^{40}$Ca, $^{56}$Ni and $^{100}$Sn, calculated 
using the Gogny D1S and the UCOM(SRG)$_{\mathrm{S,\delta 3N}}$ interactions. 
The position of the giant dipole resonance (GDR) peak, marked by arrows in the figure, is well reproduced by both interactions. 
A feature that we find with all interactions mentioned in Sec.~\ref{S:theory} is a strong ISD state at low energy, which we identify as the experimentally observed (in $^{16}$O and $^{40}$Ca) IS-LED, or isospin-forbidden $E1$ transition. 
It lies below the unperturbed spectrum (not shown), which does not contain such strong low-lying structures, and it carries very little $E1$ strength. 
The precise position and strength of the IS-LED depend on the interaction. 
Strong ISD states appear throughout the spectrum. 
At high energies, they are predominantly IS and can be identified as parts of the dipole compression mode. 
Strong IS states though appear also between the GDR and the compression mode, especially in the lighter nuclei.  
We also notice that some IS and $E1$ strength appears between the IS-LED and the GDR, and more so for larger $A$. 
By inspecting the corresponding transition densities, we find that some of these intermediate states are predominantly isovector, while others 
resemble rather IS proton-skin oscillations. 
Although these states are obviously interesting in the context of pygmy dipole, as well as IS and toroidal, strength studies, we shall refrain from further analyzing them in the present work and focus on the IS-LED modes. 

In the following we will first present the basic properties of the IS-LED. 
Then we will argue that the IS-LED is a collective, distinct mode of vibration. 
Furthermore, we will corroborate the correspondence we have made with the experimentally observed isospin-forbidden states by examining the electroexcitation cross section,  for which data exist.

\subsection{Properties and collective nature of the IS-LED} 
\label{sS:properties} 

\begin{figure*} 
\centering\includegraphics[angle=-90,width=15.0cm]{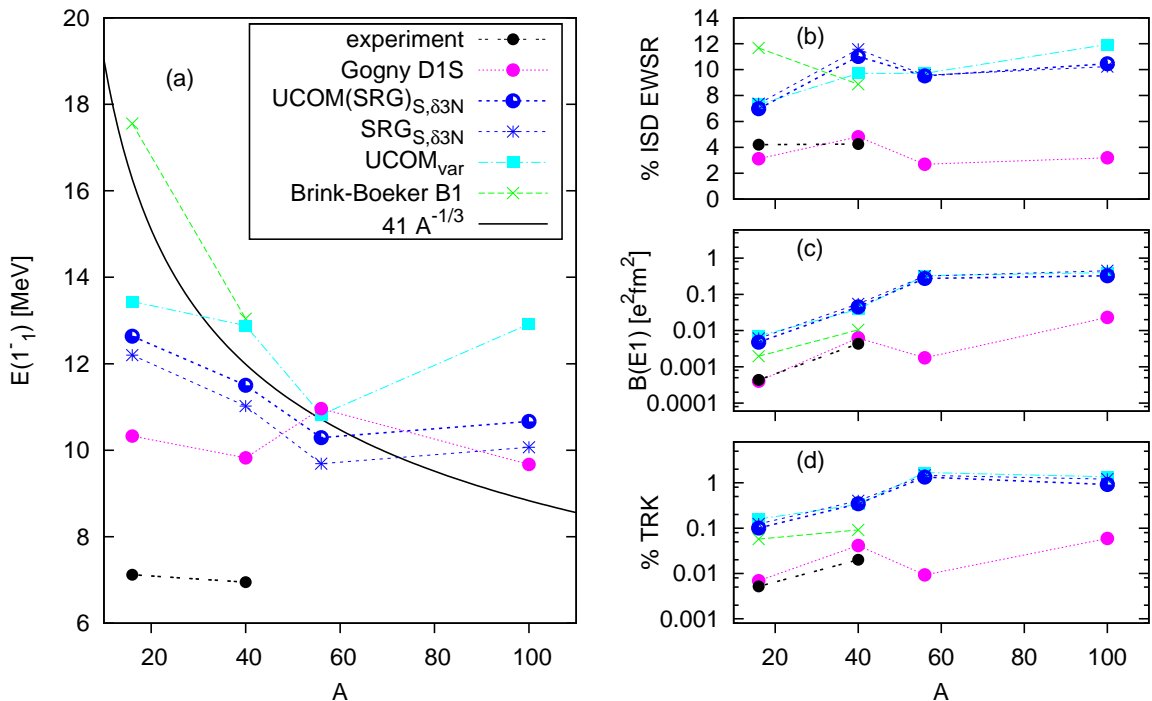} 
\caption{%
Properties of the IS-LED mode in $^{16}$O, $^{40}$Ca, $^{56}$Ni, and $^{100}$Sn as a function of mass number. 
Theoretical values correspond to only the first eigenenergy (strongly IS) obtained with various interactions (see text).  
(a) Energy; experimental values from \cite{ASe1986,CaS2004}. 
(b) Percentage of the IS EWSR; data from \cite{HaD1981,Poe1992}.  
(c) and (d) Electromagnetic excitation strength and percentage of TRK sum rule; data from \cite{ASe1986,CaS2004}.%
\label{F:vsA}}
\end{figure*}

In fig.~\ref{F:vsA} we show the basic properties of the IS-LED, calculated with the interactions Gogny D1S, UCOM(SRG)$_{\mathrm{S,\delta 3N}}$, 
SRG$_{\mathrm{S,\delta 3N}}$ and UCOM$_{\mathrm{var}}$ and how they vary with 
mass number $A$. For the $\ell -$closed nuclei $^{16}$O and $^{40}$Ca we also show results with the Brink-Boeker B1 potential, which contains no spin-orbit term. 
We show, in particular, the excitation energy, percentage of the ISD EWSR, absolute $E1$ strength, and percentage of the TRK sum carried by the IS-LED. 
Experimental data for $^{16}$O and $^{40}$Ca are included.  
Regarding the toroidal nature of the IS-LED mode, we find that it carries approximately 
$15-30$\% of the total vortical strength corresponding to the IS convection current~\cite{RaW1987}  
and $4-12$\% of its energy-weighted sum.  

The IS-LED mode carries a non-negligible amount of IS strength, 
namely $3-13\%$ of the energy-weighted strength -- depending more on the interaction used and less on the nucleus -- 
in agreement with existing experimental data for $^{16}$O and $^{40}$Ca. 
The Gogny D1S interaction overestimates the excitation energy, but otherwise agrees very well with experiment. 
All AV18-based Hamiltonians give very similar results, indicating consistency among the different unitary-transformation schemes. 
These Hamiltonians overestimate the $E1$ strength. 
They also overestimate the IS EWSR and the energy, so that the absolute IS strength may be considered in fair agreement with experiment. 
 
Both Gogny and AV18-based interactions show the same systematics for the ISD EWSR, but not for the other quantities. 
Nevertheless, they all predict that the percentage of energy-weighted $E1$ strength it carries (TRK sum) 
increases by one order of magnitude in going from $^{16}$O to $^{100}$Sn -- which means that the absolute $E1$ strength increases by two orders of magnitude.  

The properties of the IS-LED do not vary smoothly with $A$. Note that the properties of another, 
collective low-energy mode, namely the $3^-_1$, as a function of $A$ show a complicated pattern related to shell closures~\cite{KiS2002}, which would 
hardly show up in the subset of nuclei examined here. 
Let us also point out that in each case we have taken into account only the single lowest-energy and predominantly IS eigenstate, 
even if a secondary IS peak is found nearby.  
The fact that $^{16}$O and $^{40}$Ca are $\ell -$closed magic nuclei, while $^{56}$Ni and $^{100}$Sn are not, could also play a role in generating the kinks in fig.~\ref{F:vsA}.  

One notices the rather different behaviour of the results obtained with the Brink-Boeker B1 interaction, in particular as regards the energy of the IS-LED. 
With vanishing spin-orbit interaction and splittings the IS-LED energy is mainly determined by the value of $1\hbar\omega$, 
a global estimate of which, 
$41 A^{-1/3}$~MeV, is also shown in fig.~\ref{F:vsA}.  
The above observation points to a possible role played by the spin-orbit coupling in determining the properties of the IS-LED. 

The large IS strength of the IS-LED already indicates a rather collective, coherent transition. 
We first quantify the collectivity of the IS-LED using a criterion proposed in ref.~\cite{CDM2009}. 
In particular, we look at the ratio $N^{\ast}/N_{ph}$, where $N_{ph}$ is the number of $ph$ configurations available in the model space 
and $N^{\ast}$ the number of $ph$ configurations that contribute to the state in question with an amount of spectroscopic strength 
\begin{equation} 
S_{ph}=|X_{ph}|^2-|Y_{ph}|^2 \geq 1/N_{ph}, 
\end{equation} 
where $X_{ph}$, $Y_{ph}$ are the RPA transition amplitudes to the IS-LED. 
$N^{\ast}=N_{ph}$ can only be achieved for very collective states, where all configurations contribute with a statistical factor $1/N_{ph}$, 
a really exceptional occurance. 
$N^{\ast}=1$ might indicate an excitation generated by one dominant $ph$ configuration, though not conclusively. 
A larger $N^{\ast}$, though, should be a reasonable indicator of collectivity. 
We find, for example, for the $9.82$~MeV state of $^{40}$Ca using the Gogny D1S interaction, 
$N^{\ast}/N_{ph}=11/180$, to be compared with the value $15/180$ for the first GDR peak at $19.34$~MeV. 
Ten out of the $N^{\ast}=11$ configurations are $1\hbar\omega$ configurations, with a hole in the $sd$ shell and a particle in the $pf$ shell. 
In total there are $16$ $1\hbar\omega$ configurations. 
If we solve the RPA only within the restricted $1\hbar\omega$ space, we obtain 
the ratios $5/16$ for the IS-LED at $10.43$~MeV and $6/16$ for the strong GDR peak at $21.95$~MeV.  
Similar results are obtained for the other nuclei. 
We conclude that the IS-LED is a collective valence transition.

As pointed out and demonstrated in ref.~\cite{Lan2009}, it is important to look not only at the collectivity of a transition in terms of $S_{ph}$ contributions, but 
also its coherence. 
We find that several $ph$ configurations, not necessarily with large $S_{ph}$, contribute an appreciable magnitude 
of  $(X_{ph}-Y_{ph})\langle p || \hat{O}_{\mathrm{ISD}} || h\rangle $. 
The majority of those contribute with the same sign, i.e., coherently, to the effect of a large total ISD strength. 
At the same time, many $ph$ configurations contribute an appreciable amount of  
$(X_{ph}-Y_{ph})\langle p || \vec{r} || h\rangle $.  
As a rule, proton and neutron configurations with otherwise the same quantum numbers contribute with the same sign.  
The summed positive contributions, however, 
cancel out the summed negative contributions, and  thus the total amplitude of the translation operator vanishes and translational invariance is conserved. 

\subsection{Structure of the IS-LED vibration} 
\label{sS:nature}

\begin{figure} 
\centering\includegraphics[angle=0,width=8.0cm]{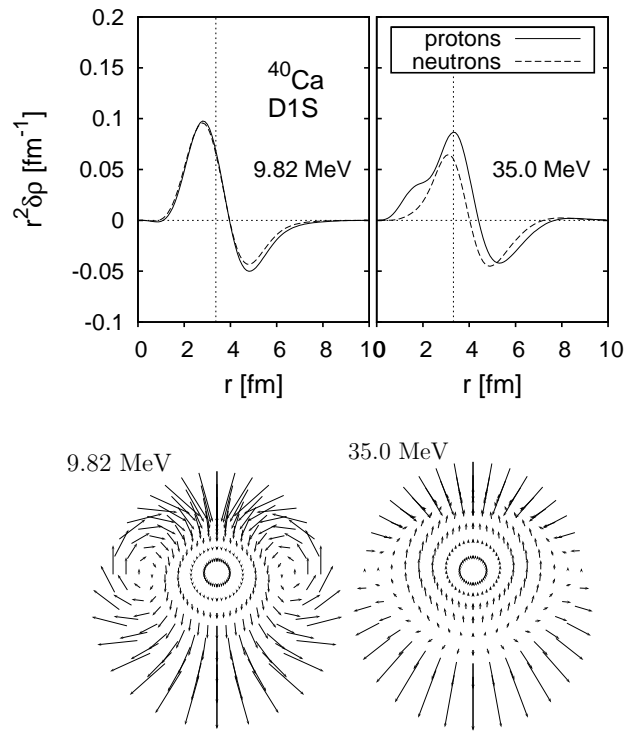} 
\caption{Proton and neutron radial transition densities 
(vertical lines mark the calculated point-nucleon mean square radius)  
and isoscalar velocity fields 
for $^{40}$Ca and the Gogny D1S interaction, for the IS-LED and the compression mode.
The distance between successive velocity vectors along each radial direction is 0.5~fm. 
} 
\label{F:drCa} 
\end{figure}

The proton and neutron transition densities, as well as the isoscalar velocity fields, 
of the IS-LED at $9.8$~MeV and the compression mode at $35.0$~MeV 
in $^{40}$Ca calculated with the Gogny D1S interaction 
are shown in fig.~\ref{F:drCa}. 
We notice that for the IS-LED the proton and neutron transition densities coincide much better than for the compression mode, making it an almost perfectly IS mode. 
Both transition densities are characterized by a node at the surface. 
The velocity field of the IS-LED does not follow the compression pattern of the high-lying state, but involves the formation of a torus 
around the surface. 
Macroscopically speaking, the IS-LED involves the translation of a core ($r<2$~fm) against a surface layer, both isoscalar and uncompressed, generating a toroidal surface oscillation. 
We find an analogous phenomenon in the IS $0\hbar\omega$ quadrupole state, which is characterized by 
the same transition density as the quadrupole giant resonance, but at the same time by vortical rather than hydrodynamical velocity fields (see, e.g., \cite{Pap2004a}). 
Let us keep in mind that, although collective, these low-lying dipole and quadrupole states 
carry only a fraction of the energy-weighted strength in their respective IS channels. As follows from the analysis in ref.~\cite{RaW1987}, 
they will therefore carry a fair amount of vortical strength.  
We have also studied the alternative toroidal operator $M_{\mathrm{tor1}}$ of ref.~\cite{KLS2003} (1st term on r.h.s. of eq.(25) there), 
essentially corresponding to out-of-phase oscillations of spin-orbit partners. We found that 
most of its strength is exhausted by a few states between the IS-LED and the GDR. Those same states 
carry very little ISD or $B(E1)$ strength.  
The IS-LED mode carries a negligible amount of $M_{\mathrm{tor1}}$ strength. 

\begin{figure} 
\includegraphics[angle=-90,width=8.0cm]{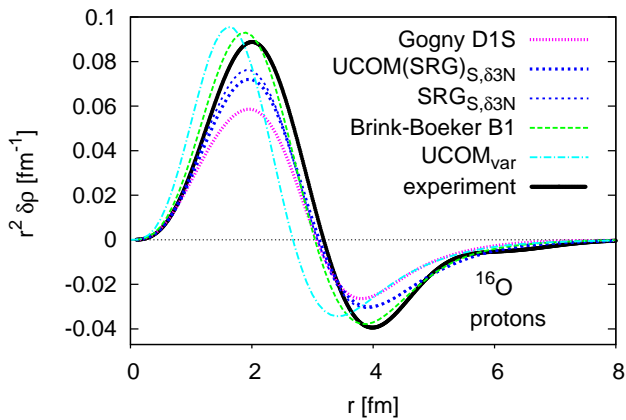} 
\caption{Proton radial transition densities for the IS-LED in $^{16}$O evaluated with different interactions. 
Experimental transition density (Laguerre expansion) taken from ref.~\cite{But1986}. 
}
\label{F:drO} 
\end{figure}

For $^{16}$O and with all but the UCOM$_{\mathrm{var}}$ interactions we find that the node 
of the proton transition density appears at $r=3.1-3.2$~fm, 
in excellent agreement with measured transition densities, \cite{But1986,COS1990}, as well as a collective model~\cite{HaD1981,COS1990}. 
Our results are shown in fig.~\ref{F:drO}. 
UCOM$_{\mathrm{var}}$ underpredicts the nuclear radius~\cite{RPP2006}, and therefore produces a node at smaller $r$. 

\begin{figure*} 
\centering\includegraphics[angle=-90,width=15.0cm]{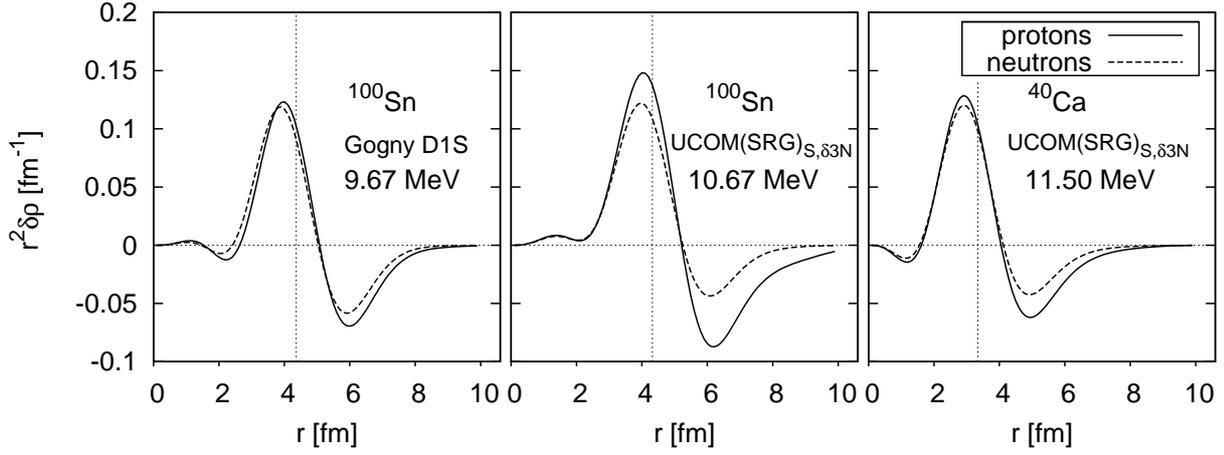} 
\caption{Proton and neutron radial transition densities for the IS-LED in $^{100}$Sn evaluated with the Gogny D1S and the UCOM(SRG)$_{S,\mathrm{\delta 3N}}$ interactions 
and in $^{40}$Ca with the UCOM(SRG)$_{S,\mathrm{\delta 3N}}$. Vertical lines mark the calculated point-nucleon mean square radius.  
} 
\label{F:drSn} 
\end{figure*}

In general, we obtain similar results regarding the nature of the IS-LED with all interactions and for all nuclei. 
There is a qualitative difference, however, between the Gogny D1S interaction, on the one hand, and the AV18-based potentials, on the other, 
leading to larger $E1$ strengths in the latter case. 
As demostrated in fig.~\ref{F:drSn} (compare UCOM(SRG)$_{\mathrm{S,\delta 3N}}$ result on $^{40}$Ca with the respective Gogny D1S results in fig.~\ref{F:drCa}), the transition densities are less perfectly isoscalar for UCOM(SRG)$_{\mathrm{S,\delta 3N}}$. Especially for larger $A$ they show an intermediate character between an IS and a proton-skin oscillation. 
Possible reasons for the discrepancy between the two types of interaction include differences in the relative importance of the Coulomb interaction and the behaviour of the symmetry energy. 
In $N=Z$ nuclei the Coulomb interaction and the symmetry energy compete against each other in the formation (Coulomb) or not (symmetry energy) of a proton skin. 
We find that the Gogny D1S interaction predicts for $^{100}$Sn a thinner proton skin 
than the AV18-derived Hamiltonians, namely 
0.82~fm or $18\%$ of the predicted charge radius,  
vs. $20\%$ of the predicted charge radius. 
Similarly, the surface dynamics and the local isospin character of the IS-LED will be determined by the interplay of Coulomb and symmetry-energy dynamics. 

By various quantitative estimates~\cite{CNV1995,SDN2009,MMS2010} 
the isospin $T=1$ admixtures in the ground state of the $T_z=0$ nuclei under study increases monotonically and smoothly with $Z$. 
According to the present results, the forbidden $E1$ strength may not provide a good absolute measure of ground-state isospin mixing, 
because it does not vary smoothly with $A$. 
Within energy-density functional theory it has been observed that isospin mixing in $N=Z$ nuclei 
does not correlate strongly with the symmetry energy or with the proton skin thickness~\cite{SDN2009,SDN2010}, in line with our elaborations.  
The problem of quantifying isospin admixtures, however, is beyond the scope of the present work.

From the form of the transition densities it becomes obvious that 
the small $B(E1)$ value of the IS-LED results from the cancellation of two large quantities: 
the $r-$weighted integral of $r^2\delta\rho_{p}(r)$ up to the node, 
$I_-$,  and the integral above the node, $I_+$. 
As a result, deviations in the transition densities, such as presented in figs.~\ref{F:drO},~\ref{F:drSn}, can lead to disproportionally large differences in the $E1$ strength. 
By setting $V_{\mathrm{Coul}}$ equal to zero we obtain a perfect cancellation and vanishing $E1$ strength, 
but the transition is no less collective for this reason. 
Its energy is almost the same as before, and its IS strength remains large.  
It is instructive to consider the following difference between the electromagnetic strength of a dipole mode and a mode of some other multipolarity, 
e.g. $J^{\pi}=3^-$. 
Assume equal proton and neutron radial transition densities throughout,  $\delta\rho_{p}(r)=\delta\rho_{n}(r)$,  
and therefore perfectly IS modes. 
In the octupole case $r^2\delta\rho_p (r)$ is surface-peaked and its $r^3-$weighted integral, yielding its $B(E3)$, will be large if its amplitude is large. 
If the mode is strong in the IS channel, it will be strong in the $E3$ channel. 
In the dipole case, however, the $r-$weighted integral, giving the $B(E1)$ strength, will be zero due to the translational-invariance condition, 
\[ 
\int_0^{\infty} \delta\rho_p(r) r^3 dr 
= 
\frac{1}{2}\int_0^{\infty} [\delta\rho_p(r)+\delta\rho_n(r)] r^3 dr 
= 0 
,\] 
even if the mode exhausts most of the IS EWSR.  
In both cases the isovector strength will be zero. 
The non-zero value of $B(E1)$ simply means that the state is not perfectly isoscalar. 

\begin{figure*} 
\centering\includegraphics[angle=0,height=9cm,width=15.0cm]{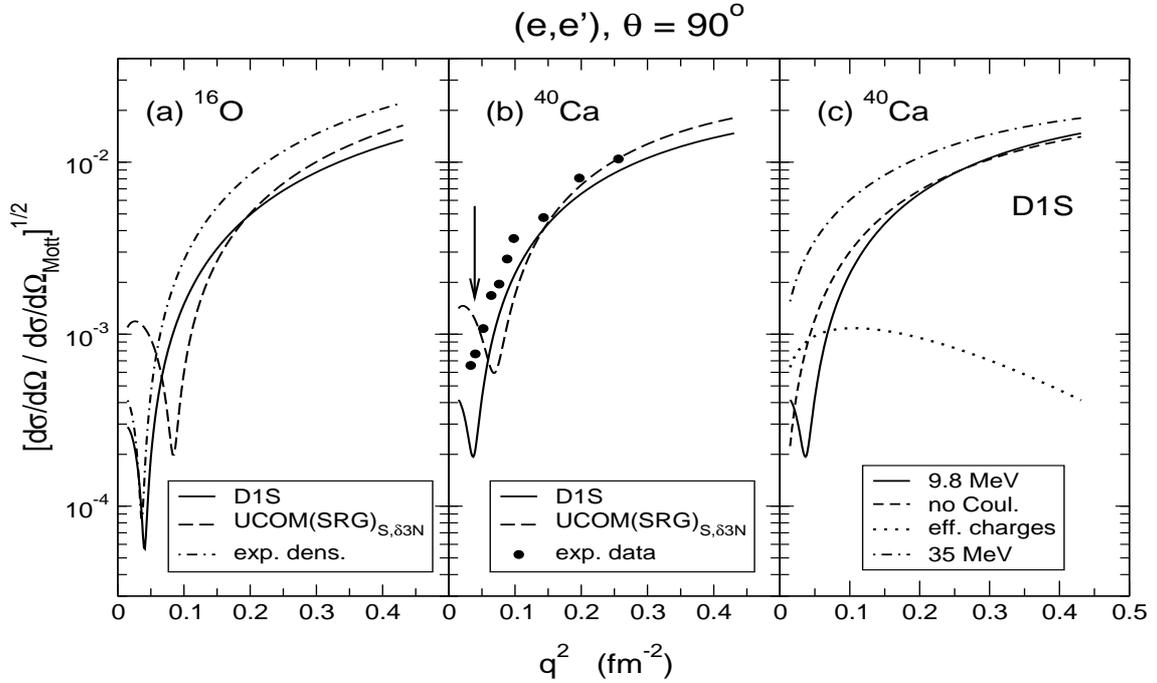} 
\caption{Electroexcitation form factor (absolute value) derived within DWBA as the square root of the cross section divided by the Mott cross section, 
for $^{16}$O and $^{40}$Ca, using the interactions Gogny D1S and UCOM(SRG)$_{S,\mathrm{\delta 3N}}$.  
(a) Calculations for the IS-LED of $^{16}$O compared with results extracted from the experimental transition density~\cite{But1986}. 
(b) Calculations for the IS-LED of $^{40}$Ca compared with data~\cite{Gra1977,Gra1977a}. 
The minimum of the reported best fit to the data, at $q_{\min}^2=0.039$~fm$^{-2}$, is indicated by an arrow. 
(c) For the D1S interaction and $^{40}$Ca, the result (b) for the IS-LED is compared: with the result 
when no Coulomb interaction is considered; when effective charges, eq.~(\ref{E:effch}), are used; and with the form factor of the peak at 35~MeV (compression mode). 
} 
\label{F:ff} 
\end{figure*}

In fig.~\ref{F:ff} we show the electroexcitation longitudinal form factor of the IS-LED in $^{16}$O and $^{40}$Ca 
within distorted-wave Born approximation (DWBA), 
to be compared with the experimental measurements~\cite{Gra1977,HeS1976,But1986}. 
The characteristic minimum is reproduced in both cases, and best by Gogny D1S. 
The form factor of the compression mode is also shown, for $^{40}$Ca with the Gogny force. 
The difference is obvious. 
Close to the photon point, and for the integrals weighted with a Bessel function $j_1(qr)$, rather than $r$, within PWBA, 
we have $|I_+|>|I_-|$ for the IS-LED. For larger $q$, 
$I_-$ gains relative strength and eventually dominates, leading to a node in the form factor and the minimum in its absolute value. 
By contrast, for the compression mode $I_-$ dominates already towards the long-wavelength limit, hence there is no minimum in the form factor. 
Thus similar, at first glance,  transition densities, characterized by a node, lead to different, even qualitatively, form factors. 
 
In fig.~\ref{F:ff}  we show also what happens if we calculate the form factor with effective charges, i.e., using the transition density  
of eq.~(\ref{E:effch}) 
instead of simply $\delta\rho_p(r)$, eq.~(\ref{E:noeffch}). 
The curves 
seriously diverge. 
Notice, finally, that even if $V_{\mathrm{Coul}}=0$ and $\delta\rho_p(r)=\delta\rho_n(r)$, 
the form factor of the IS-LED state is not zero for $q>0$, contrary to what would be obtained with the use of effective charges. 

\section{Discussion} 
\label{S:disc} 

We have found that the IS-LED mode is a collective, coherent transition  
and that its energy may depend on the spin-orbit coupling. 
The $E1$ strength it carries is a rather delicate matter. 
As we saw, it is due to  
the Coulomb interaction -- for charge-symmetric $V_{\mathrm{NN}}$ -- whose role in breaking the local isospin symmetry increases with $Z=A/2$, 
leading, for example, to the formation of a thicker proton skin~\cite{Aue2010}. 
There appears to be some correlation, within RPA, between the amount of IS and $E1$ strength predicted by different interactions. 
We have argued that the precise $E1$ strength carried by the IS-LED is determined by the interplay of the Coulomb interaction and the symmetry energy 
or its slope. 
The latter affects also the properties of the GDR. 
There is a good chance then that an RPA or other model reproducing correctly the properties of the IS-LED state 
and at the same time those of the GDR, 
should be able to describe satisfactorily the $E1$ strength and other low-energy dipole phenomena as well -- a speculation worth investigating in the future. 
As far as self-consistent HF-RPA is concerned, the above demand is not trivial, 
as the quality of an RPA description of low-lying states and giant resonances depends in different ways on the properties of the single-particle spectrum. 
Effects beyond RPA may be indispensible. 

A related question is whether the IS-LED mode has a counterpart in other nuclei, notably $N>Z$ ones, where a neutron skin may develop,  
that could oscillate against the isospin-saturated core. 
It cannot be ruled out that the core can undergo its own inner-core--vs--layer excitation, be it Pauli-suppressed due to the additional occupied neutron levels. 
Such a mode, if it develops, would likely carry little $E1$ strength, but feature prominently in the ISD strength function. 
Moreover, it could mix with a possible neutron-skin oscillation and influence its position and strength. 
The structure of the low-energy dipole spectrum would be richer than hitherto predicted. 
It is tempting to regard a neutron-skin mode as a special case of an IS-LED mode. 
To the extent that they both are coherent $1\hbar\omega$ states this is a valid classification.  
However, the proton transition density is very different in the two cases 
and an ideal neutron-skin oscillation, with a nodeless proton transition density, 
would carry much more $E1$ strength. 
In any case, an electron scattering experiment would be able to establish the character of a given state. 
The above issues will be the subject of future work. 

Up to now we have focused on a self-consistent non-relativistic RPA with finite-range forces, both phenomenological and realistic. 
Next we ask whether other models give similar results. 
By inspecting the literature on isospin-forbidden $E1$ strength we find no evidence that the conventional shell model 
produces $1^-$ collectivity in the $1\hbar\omega$ 
regime, even though a $T=0$ dipole state with the correct $B(E1)$ value can be obtained with suitable adjustments. 
Similarly, we cannot conclude on the RPA results reported in refs.~\cite{Har2004,Ter2007}, based on empirical single-particle states. 
We found that a simple RPA model with single-particle states generated by a Woods-Saxon potential and a separable dipole-dipole residual interaction, 
with adjustments for the energy of the spurious mode to vanish, 
does not produce $1\hbar\omega$ collectivity. 

Regarding other self-consistent $ph$ models, to the best of our knowledge there are no reports of the IS-LED mode in $N=Z$ nuclei within 
modern relativistic RPA. 
Some Skyrme-RPA results are tabulated in 
\cite{Hey1990} and were first reported in \cite{War1982,WWH1983}. 
The SkE4* Skyrme parameterization is used, which includes an additional three-body term.  
The position and $B(E1)$ strengths of the $T=0$ states are fairly well reproduced for both $^{16}$O and $^{40}$Ca. 
Energetically they appear almost 2~MeV below the first unperturbed $ph$ configuration, therefore they could be collective, 
but no information on their IS strenght is provided. 
Similarly, it is difficult to conclude on the Skyrme-RPA results of ref.~\cite{BlG1977} and those of an early application of relativistic field theory~\cite{Fur1985}. 

Systematic self-consistent RPA calculations, employing standard Skyrme functionals, have been reported for the ISD strength function of Ca, Ni and Sn isotopes~\cite{TeE2006}. 
A rather strong IS-LED mode is apparently predicted for the $N=Z$ isotopes, though it is not further analyzed as such. 
For $^{40}$Ca its energy is fairly close to the observed IS-LED. 
The lowest ISD state of $^{56}$Ni is not as prominent as that of $^{40}$Ca and $^{100}$Sn, 
not unlike our prediction 
with the Gogny D1S interaction, see fig.~\ref{F:response}.  
The isovector dipole EWSR below 10~MeV is overestimated in $^{40}$Ca by a factor of $4.5$. 

Using a fairly self-consistent RPA model (spin-dependent and Coulomb terms missing from the residual interaction but no restrictions imposed on the $ph$ space) 
and the parameterizations SkM* and BSk8 as representatives of the Skyrme species 
we calculated the ISD response of $^{16}$O and $^{40}$Ca. 
We found two very weak states below 10~MeV which correspond, within 0.5~MeV ($^{16}$O) or 0.1~MeV ($^{40}$Ca), to unperturbed $ph$ excitations of comparable $E1$ strength. 
Thus they can hardly be considered collective or account for the experimentally observed ISD strength, 
at least for $^{40}$Ca. 
However, by omitting the spin-orbit term of the residual interaction in this calculation we may have missed an important effect~\cite{BlG1977}.  
In view of the results of ref.~\cite{TeE2006} 
as well as investigations of inconsistency effects~\cite{ASS2003}, the degree of consistency must be very important when studying the IS-LED within RPA. 

Finite-range interactions have not been extensively used to describe closed-shell nuclei. 
An early RPA application of the Gogny D1 interaction focused on high-lying excitations and neglected states exhausting less than 
4\% of the ISD EWSR~\cite{DeS1983} -- approximately the amount of strength carried by the IS-LED. 
In ref.~\cite{BlG1977} the energy but not the strength of the IS-LED of $^{16}$O was calculated using Gogny and Skyrme forces. 
The importance of a correct spin-orbit splitting for reproducing the experimental energy of this state was stressed. 

It is worth mentioning that a core-layer dipole nuclear vibration has been studied macroscopically~\cite{Bas2008}, 
but it is difficult to establish a quantitative connection with our work as the focus there was on the pygmy dipole strength of $N>Z$ nuclei. 
Last but not least, a strong ISD mode at the correct energy and with a similar transition density as we find has been 
tentatively reported for $^{40}$Ca within the Extended Theory of Finite Fermi Systems~\cite{Ter2006}. 

\section{Summary and outlook} 
\label{S:concl}

In summary, we have found that the self-consistent RPA with finite-range interactions, 
such as Gogny parameterizations and transformed Hamiltonians based on realistic potentials, 
predict the existence of an isoscalar, low-energy dipole (IS-LED) mode in spherical, closed-shell nuclei, 
in agreement with available experimental data. 
Here we have focused on $N=Z$ nuclei, but   
rather strong low-lying IS states have been detected also in heavier stable nuclei~\cite{HaV2001s1,Poe1992}, 
below the alleged toroidal dipole mode. 
The IS-LED involves the oscillation of an uncompressed surface layer against a core. 
Although certain types of calculations appear to not even predict the existence of such a mode, 
available experimental data corroborate our results.
In particular, the strength and electroexcitation form factor of the apparently collective 
low-energy ISD state of $^{16}$O and $^{40}$Ca can be simultaneously accounted for. 
The presence and small amount of $E1$ strength are related, respectively, 
to the Coulomb interaction and the form of the transition densities, dictated by translational invariance. 
The excitation energy is systematically overestimated. 
The percentage of TRK sum carried by the IS-LED may grow by one order of magnitude when going from the lightest ($^{16}$O) to the heaviest nucleus studied ($^{100}$Sn) and the absolute $E1$ strength by two orders of magnitude. 
The percentage of IS strength changes more moderately.  

The precise trend of the energy, strength and fragmentation as we go to heavier $N=Z$ depends on the interaction used. 
Of course, the effective Hamiltonians used are not tailored for the present delicate application. 
As possible relevant properties we identify the symmetry energy and its interplay with the Coulomb potential, as well as the spin-orbit coupling. 
We have speculated that a microscopic model, not necessarily RPA, which correctly reproduces the energy and IS strength of the IS-LED mode, 
along with the properties of the GDR, 
could lead to a reliable description of the $E1$ strength and other low-energy dipole phenomena. 
Implications for the pygmy dipole strength of $N\neq Z$ nuclei and its isospin structure will be the subject of future work.

\section*{Acknowledgments} 
We wish to thank Profs. A. Richter and P. von Neumann-Cosel for pertinent information and useful remarks. 
This work was supported by the Deutsche Forschungsgemeinschaft 
through SFB 634, by the Helmholtz International Center for FAIR 
within the framework of the LOEWE program launched by the state of Hesse, 
and by the BMBF Verbundforschung (Contract 06DA90401).

%
%


\end{document}